\documentclass[prb,showpacs,twocolumn]{revtex4}

\usepackage{graphicx}
\usepackage{amssymb}

\def\crea{^{\dagger}}
\def\anni{^{\phantom{\dagger}}}

\newcommand{\rd}{\mathrm{d}}
\newcommand{\re}{\mathrm{e}}
\newcommand{\ri}{\mathrm{i}}
\newcommand{\Imm}{\,\mathrm{Im}\,}

\newcommand{\const}{\mathrm{const}}

\newcommand{\reff}{\mathrm{eff}}
\newcommand{\req}{Eq.~}
\newcommand{\fig}{Fig.~}

\newcommand{\rl}{\langle\langle}
\newcommand{\rr}{\rangle\rangle}

\begin{document}

\title{Densities of states of the Falicov-Kimball model off half filling in infinite dimensions}

\author{Ihor~V.~Stasyuk}
\author{Orest~B.~Hera}
\email{hera@icmp.lviv.ua}
\affiliation{%
Institute for Condensed Matter Physics of the National Academy of
Sciences of Ukraine,\\ 1~Svientsitskii Str., 79011 Lviv, Ukraine}

%\date{\today}
\date{February 28, 2005}
%%\date{Received February 28, 2005}

\begin{abstract}
An approximate analytical scheme of the dynamical mean field
theory (DMFT) is developed for the description of the electron
(ion) lattice systems with Hubbard correlations within the
asymmetric Hubbard model where the chemical potentials and
electron transfer parameters depend on an electron spin (a sort of
ions). Considering a complexity of the problem we test the
approximation in the limiting case of the infinite-$U$ spinless
Falicov-Kimball model. Despite the fact that the Falicov-Kimball
model can be solved exactly within DMFT, the densities of states
of localized particles have not been completely investigated off
half filling. We use the approximation to obtain the spectra of
localized particles for various particle concentrations (chemical
potentials) and temperatures. The effect of a phase separation
phenomenon on the spectral function is considered.
\end{abstract}


\maketitle

\section{Introduction}

The Falicov-Kimball model \cite{Fal69} was introduced to describe
the thermodynamics of metal-insulator transitions in  compounds
that contained both itinerant and localized quasiparticles. The
spinless Falicov-Kimball model is the simplest example of an
interacting fermionic system that displays numerous phase
transitions. Despite its relative simplicity, the analysis of the
model is very complex and still a lot of open problems remain. The
model can describe many phenomena such as metal-insulator
transition, ferromagnetism, antiferromagnetism, phase separations,
etc.

In the dynamical mean-field theory \cite{Met89,Geo96} which is
exact in infinite dimensions \cite{Mul89} the Falicov-Kimball
model can be solved exactly by calculating the grand canonical
potential and the single-particle Green's function
\cite{Bra89,Bra90,Bra91}. The thermodynamics of the
Falicov-Kimball model is well investigated. The phase transitions
between homogeneous phases or phase separations are investigated
using the mentioned exact methods \cite{Sta03,Fre99} as well as
using the effective Hamiltonian at the large-$U$ limit
\cite{Let99} (see Ref.~\onlinecite{Fre03} for detailed review of
the Falicov-Kimball model).

The problem of an evaluation of the spectral function of localized
particles is more complex than the investigation of thermodynamics
and the phase transitions. The $f$ spectral function related to
the localized particles was calculated at half-filling more than
ten years ago \cite{Bra92}. Recently, the exact scheme has been
extended to the case of different particle concentrations
\cite{Zla01}, but due to computational difficulties the densities
of states of localized particles have not been completely
investigated off half filling. So, we suggest here an approximate
analytical scheme within the dynamical mean-field theory for
calculating the $f$ spectral function at different particle
concentrations and temperatures.

We consider the asymmetric Hubbard model describing the dynamics
of two types of particles (ions, electrons or quasiparticles) as a
generalization of the Falicov-Kimball model. The Hamiltonian of
the asymmetric Hubbard model in a second quantization has the
following form
\begin{equation}
H=-\sum_{i\sigma} \mu_{\sigma} n_{i\sigma} + U \sum_i
n_{i\uparrow}n_{i\downarrow} + \sum_{ij\sigma} t_{ij}^{\sigma}
a_{i\sigma}\crea a_{j\sigma}\anni\,,
\end{equation}
where $n_{i\sigma}=a_{i\sigma}\crea a_{i\sigma}\anni$ and the
motion of particles is described by the creation
($a_{i\sigma}\crea$) and annihilation ($a_{i\sigma}\anni$)
operators. The chemical potentials $\mu_{\sigma}$ and the transfer
parameters $t_{ij}^{\sigma}$ depend on a sort of particles (an
electron spin). The value $U$ describes the local on-site
repulsion.

The asymmetric Hubbard model was proposed for the description of
mixed-valence compounds \cite{Koc94}. This model can also be used
for the investigation of the lattice systems having an ionic
conductivity with two types of ions. When a single lattice site
can be occupied only by one ion, the nontrivial limit
$U\rightarrow\infty$ should be considered. Is this case, various
thermodynamic regimes can be realized. The chemical potentials or
concentrations of particles of different sorts can be fixed
independently. In this context the model can be investigated in
the presence of the external field corresponding to the difference
between the chemical potentials of different sorts.

A number of methods for describing the strongly correlated
electron systems has been developed within the dynamical mean
field theory. However, all these methods have various
restrictions. The quantum Monte-Carlo method
\cite{Jar92,Roz92,Geo92} is numerically exact but has severe
problems at low temperatures and for high repulsion strength $U$.
The exact diagonalization method \cite{Caf94,Si94} is restricted
to a small number of orbitals. Among the numerical techniques the
most reliable one at low temperatures is the numerical
renormalization group method \cite{Bul00}. For example its
extension was used for the description of the ground state of the
standard Hubbard model \cite{Zit02}.

Besides the numerical approaches the development of analytical
approximations for the infinite-dimensional model still remains
necessary. The Hubbard model in the large-$U$ limit was
investigated using the non-crossing approximation
\cite{Pru93,Obe97}. Many approximations were developed for the
weak-coupling regime, for example, the Edwards-Hertz approach
\cite{Edw90,Edw93,Wer94,Wer95}. The alloy-analogy based
approximations \cite{Her96,Pot98,Pot98b} do not take into account
the effect of scattering processes on forming the energy band and
cannot be used for the investigation of spectra of the asymmetric
Hubbard model. In the Falicov-Kimball limit, the alloy-analogy
(AA) and modified alloy-analogy (MAA) approximations give the
density of states of localized particles in the form of a delta
function. The scattering processes should be taken into account
for a correct description of the broadening of this peak. For
example, it was the Hubbard-III approximation \cite{Hub64b} that
originally included the electron scattering for the half-filled
Hubbard model into the theory.

We use and improve the approximate analytical approach originally
proposed for the Hubbard model \cite{Sta00} and extended to the
asymmetric Hubbard model \cite{Sta03}. In this method the
single-site problem is formulated in terms of the auxiliary
Fermi-field. The approach is based on the equations of motion and
on the irreducible Green's function technique with projecting on
the basis of Fermi-operators. This approach gives DMFT equations
in the approximation which is a generalization of Hubbard-III
approximation and includes as simple specific cases the AA and MAA
approximations.

The approximation is tested on the infinite-$U$ spinless
Falicov-Kimball model. We use the approximation to obtain the
densities of states of localized particles for various chemical
potentials (concentrations) and temperatures. The dependences of
the chemical potentials on the particle concentrations are
calculated using the densities of states as well as
thermodynamically by calculating the grand canonical potential
\cite{Sta02}. To calculate the particle spectra at low
temperatures the phase separation should be taken into account.

In Section~II, we review the formalism of DMFT with the use of the
auxiliary Fermi-field and the approximate analytical scheme based
on the projecting technique and the different-time decoupling
procedure. In Section~III and in Appendix, the exact relations
between some Green's functions are derived, which allows us to
find the projecting coefficients using only the single particle
Green's function and the coherent potential. The results are
discussed in Section~IV, followed by our conclusions in Section~V.

\section{Formalism}

In the dynamical mean-field theory the infinite-di\-men\-sion\-al
lattice model is mapped on the single-site problem
{\arraycolsep=-0.5ex
\begin{eqnarray}
&&{} \re^{-\beta H} \rightarrow \re^{-\beta H_{\reff}} =
\re^{-\beta H_0} \nonumber\\
&&{} \times \mathcal{T}\exp \Bigg[ - \int_0^{\beta} \!\! \rd\tau
\!\! \int_0^{\beta} \!\!\rd\tau' \! \sum_{\sigma}
J_{\sigma}(\tau-\tau') a_{\sigma}\crea (\tau) a_{\sigma}\anni
(\tau') \Bigg]
\end{eqnarray}}%
with the coherent potential $J_{\sigma}(\tau-\tau')$ which has to
be self-consistently determined from the conditions
\begin{eqnarray}
&&G_\sigma(\omega_n)=\frac{1}{\Xi_\sigma^{-1}(\omega_n)-J_\sigma(\omega_n)},
 \label{sys2} \\
&&G_\sigma (\omega_n,
\mathbf{k})=\frac{1}{\Xi_\sigma^{-1}(\omega_n)-t^\sigma_\mathbf{k}},
  \label{sys1} \\
&&G_\sigma(\omega_n)=G_{ii}^\sigma(\omega_n)=\frac{1}{N}\sum_\mathbf{k}
G^\sigma
  (\omega_n, \mathbf{k}),
  \label{sys3}
\end{eqnarray}%
where $G_{\sigma}$ is the one-particle Green's function,
$\Xi_{\sigma}$ is the total irreducible part which does not depend
on the wave vector $\mathbf{k}$. The sum over $\mathbf{k}$ in
\req\ref{sys3} is calculated by the integration with the density
of states (DOS) (a Gaussian DOS for an infinite-dimensional
hypercubic lattice and a semielliptic DOS for a $d=\infty$ Bethe
lattice).

 It was shown in Ref.~\onlinecite{Sta00} that the single-site problem can be formulated
in terms of the auxiliary Fermi-operators $\xi_{\sigma}$
describing the creation and the annihilation of particles in the
effective environment. The problem is described by the following
Hamiltonian
\begin{equation}
H_{\reff}=H_0+\sum_{\sigma} V_{\sigma}\Big(a_{\sigma}\crea
\xi_{\sigma}\anni\! +\xi_{\sigma}\crea a_{\sigma}\anni
\!\Big)+H_{\xi}.
\end{equation}
The approach does not require an explicit form of the environment
Hamiltonian $H_{\xi}$. The environment is described by the
coherent potential given as the Green's function for the auxiliary
Fermi-field with the unperturbed Hamiltonian $H_\xi$:
\begin{equation}
J_{\sigma}(\omega)=2\pi V^2_{\sigma}
\langle\langle\xi_{\sigma}\anni | \xi_{\sigma}\crea
\rangle\rangle_{\omega}^{\xi}.
\end{equation}

The particle creation and annihilation operators are expressed in
terms of Hubbard operators
\begin{equation}
a_{\sigma}= X^{0\sigma} + \zeta X^{\bar{\sigma} 2}
\end{equation}
on the basis of single-site states $| n_A, n_B\rangle$
\begin{equation}
\begin{array}{lll}
|0\rangle=|0,0\rangle , &  &\qquad |A\rangle=|1,0\rangle, \\
|2\rangle=|1,1\rangle, &  &\qquad |B\rangle=|0,1\rangle ,
\end{array}
\label{BaseStates2}
\end{equation}
where the following notations for sort indices are used:
$\bar{\sigma}=B$, $\zeta=+$ for $\sigma=A$ and $\bar{\sigma}=A$,
$\zeta=-$ for $\sigma=B$. In this representation the local
Hamiltonian $H_0$ of the asymmetric Hubbard model is
\begin{equation}
  H_{0}  =  - \sum_{\sigma} \big[  \mu_{\sigma} \big( X^{\sigma\sigma}+X^{22}
  \big) \big] + U X^{22}\,,
\end{equation}
and the two-time Green's function $G_\sigma(\omega) \equiv 2\pi
\langle\langle a_\sigma\anni | a_\sigma\crea
\rangle\rangle_\omega$ is written as:
{\arraycolsep=2pt
\begin{eqnarray}
 G_{\sigma} & = & 2\pi\big[\langle\langle X^{0\sigma}| X^{\sigma 0}
\rangle\rangle_{\omega}
 +\zeta \langle\langle X^{0\sigma}| X^{2\bar{\sigma}}
 \rangle\rangle_{\omega}  \nonumber \\
& &{} +\zeta \langle\langle X^{\bar{\sigma}2}| X^{\sigma0}
\rangle\rangle_{\omega}
 +\langle\langle X^{\bar{\sigma}2}| X^{2\bar{\sigma}}
 \rangle\rangle_{\omega}\big]\,.
 \label{Grin_X}
\end{eqnarray}}%

The Green's functions in \req\ref{Grin_X} are calculated using the
equations of motion for Hubbard operators:
\begin{eqnarray}
\ri\frac{\rd}{\rd t} X^{0\sigma
(\bar{\sigma}2)}(t)=\big[X^{0\sigma (\bar{\sigma}2)}\,,
H_{\reff}\big]\,. \label{eq_motion01}
\end{eqnarray}%
The commutators (\ref{eq_motion01}) are projected on the subspace
formed by operators $X^{0\sigma}$ and $X^{\bar{\sigma}2}$:
\begin{equation}
\big[X^{\gamma}, H_{\reff}\big]=\alpha_1^{\gamma} X^{0\sigma}+
\alpha_2^{\gamma} X^{\bar\sigma 2} + Z^{\gamma}\,.
\end{equation}
The operators $Z^{0\sigma (\bar{\sigma}2)}$ are defined as
orthogonal to the operators from the basic subspace
\cite{Sta00,Sta02,Sta03}:
\begin{equation}
\langle \{Z^{0\sigma (\bar{\sigma}2)}, X^{0\sigma
(\bar{\sigma}2)}\}\rangle = 0.
\end{equation}
These equations determine the projecting coefficients
$\alpha_{i}^{0\sigma (\bar{\sigma}2)}$ which are expressed in
terms of the mean value
\begin{equation}
 \varphi_\sigma =
 \langle \xi_{\bar\sigma}\anni X^{\bar\sigma 0}\rangle +
 \zeta \langle X^{\sigma 2} \xi_{\bar\sigma}\crea \rangle.
\end{equation}

Using this procedure by differentiating both with respect to the
left and to the right time arguments, we come to the relations
between the components of the Green's function $G_\sigma$ and
scattering matrix $\hat{P}_\sigma$. In a matrix representation, we
have
\begin{equation}
 \hat{G}_\sigma=\hat{G}_0^\sigma+\hat{G}_0^\sigma\hat{P}_\sigma\hat{G}_0^\sigma,
\end{equation}
where
\begin{equation}
  \hat{G}_\sigma=2\pi \left(
  \begin{array}{lll}
  \langle\langle X^{0\sigma}| X^{\sigma 0} \rangle\rangle & &
  \langle\langle X^{0\sigma}| X^{2 \bar\sigma} \rangle\rangle \\
  \langle\langle X^{\bar\sigma 2}| X^{\sigma 0} \rangle\rangle & &
  \langle\langle X^{\bar\sigma 2}| X^{2 \bar\sigma} \rangle\rangle
  \end{array}
  \right),
  \label{Grin_Mart}
\end{equation}
and nonperturbed Green's function $\hat {G}^\sigma_0$ is
{\arraycolsep=1.6pt
\begin{equation}
  \hat{G}^\sigma_0=\frac{1}{D_\sigma} \left(
  \begin{array}{ccc}
  \omega-b_\sigma & & -\zeta\frac{V_\sigma}{A_{2\bar\sigma}}\varphi_\sigma \\
  -\zeta\frac{V_\sigma}{A_{0\sigma}}\varphi_\sigma & & \omega-a_\sigma
  \end{array}
  \right) \left(
  \begin{array}{ccc}
  A_{0\sigma} & & 0 \\
  0  & & A_{2\bar\sigma}
  \end{array}
  \right),
\end{equation}}
where
\begin{equation}
A_{pq}= \langle X^{pp}+X^{qq}\rangle,\,\,
A_{0\sigma}=1-n_{\bar\sigma},\,\, A_{2\bar\sigma}=n_{\bar\sigma},
\end{equation}
\begin{equation}
  D_\sigma=(\omega-a_\sigma)(\omega-b_\sigma)-
  \frac{V^2_\sigma}{A_{0\sigma}A_{2\bar\sigma}}\varphi_\sigma^2,
\end{equation}
\begin{equation}
  a_\sigma=-\mu_\sigma+\frac{V_\sigma}{A_{0\sigma}}\varphi_\sigma,
  \quad
  b_\sigma=U-\mu_\sigma+\frac{V_\sigma}{A_{2\bar\sigma}}\varphi_\sigma.
\end{equation}
The scattering matrix
{\arraycolsep=-5pt
\begin{eqnarray}
&&{}  \hat{P}_\sigma=2\pi
  \left(
  \arraycolsep=2pt
  \begin{array}{ccc}
  A_{0\sigma}^{-1} & & 0 \\
  0  & & A_{2\bar\sigma}^{-1}
  \end{array}
  \right) \nonumber\\
&&{}  \times \left(
  \arraycolsep=2pt
  \begin{array}{ccc}
  \langle\langle Z^{0\sigma}| Z^{\sigma 0} \rangle\rangle & &
  \langle\langle Z^{0\sigma}| Z^{2 \bar\sigma} \rangle\rangle \\
  \langle\langle Z^{\bar\sigma 2}| Z^{\sigma 0} \rangle\rangle & &
  \langle\langle Z^{\bar\sigma 2}| Z^{2 \bar\sigma} \rangle\rangle
  \end{array}
  \right)
  \left(
  \arraycolsep=2pt
  \begin{array}{ccc}
  A_{0\sigma}^{-1} & & 0 \\
  0  & & A_{2\bar\sigma}^{-1}
  \end{array}
  \right)
  \label{scattmatr}
\end{eqnarray}}%
being expressed in terms of irreducible Green's functions contains
the scattering corrections of the second and the higher orders in
powers of $V_\sigma$. The separation of the irreducible parts in
$\hat{P}_\sigma$ enables us to obtain the mass operator
$\hat{M}_\sigma=\hat{P}_\sigma|_{\mathrm{ir}}$ and the single-site
Green's function expressed as a solution of the Dyson equation
\begin{equation}
 \hat{G}_\sigma=(1-\hat{G}^\sigma_0\hat{M}_\sigma)^{-1} \hat{G}^\sigma_0.
 \label{dyson1}
\end{equation}

We restrict ourselves to the simple approximation in calculating
the mass operator $\hat {P}_\sigma$, taking into account the
scattering processes of the second order in $V_\sigma$. In this
case
\begin{equation}
 \hat{M}_{\sigma}= \hat{P}_{\sigma}^{(0)},
 \label{massop}
\end{equation}
where the irreducible Green's functions are calculated without
allowance for correlation between electron transition on the given
site and environment. It corresponds to the procedure of
different-time decoupling, which means in our case an independent
averaging of the products of $X$ and $\xi$ operators.

Let us illustrate this approximation on the example of calculation
of the following irreducible Green's function:
\begin{equation}
I (\omega) \equiv \langle\langle
{(X^{00}+X^{\sigma\sigma})\xi_\sigma\anni} |
{\xi_\sigma\crea(X^{00}+X^{\sigma\sigma})}
\rangle\rangle_\omega^{\mathrm{ir}}\,.
\end{equation}
According to the spectral theorem, this Green's function is
related to the corresponding time correlation function, and
according to the different-time decoupling we have
{\arraycolsep=-2pt
\begin{eqnarray}
&&{} \langle
  \xi_\sigma\crea(t)(X^{00}+X^{\sigma\sigma})_t
  (X^{00}+X^{\sigma\sigma})\xi_\sigma\anni
 \rangle^{\mathrm{ir}} \nonumber\\
&& {\qquad} \approx \langle (X^{00}+X^{\sigma\sigma})_t
 (X^{00}+X^{\sigma\sigma}) \rangle
 \langle
  \xi_\sigma\crea (t) \xi_\sigma\anni
 \rangle.
 \label{e43}
\end{eqnarray}}%
Calculation of these correlation functions in a zero approximation
{\arraycolsep=-2pt
\begin{eqnarray}
&&{} \langle (X^{00}+X^{\sigma\sigma})_t
(X^{00}+X^{\sigma\sigma}) \rangle \nonumber\\
&&{\qquad}  \approx \langle (X^{00}+X^{\sigma\sigma})^2 \rangle =
A_{0\sigma}
\end{eqnarray}}%
leads to the result
\begin{equation}
  I(\omega)=A_{0\sigma}\langle\langle \xi_\sigma\anni |
  \xi_\sigma\crea
  \rangle\rangle_\omega^\xi =
  \frac{A_{0\sigma}}{2\pi V^2} J_\sigma (\omega).
\end{equation}

Using the above procedure we can obtain the final expressions for
the mass operator and the total irreducible part:
 {\arraycolsep=0.5pt
\begin{eqnarray}
 \Xi_\sigma^{-1}(\omega)&=&\bigg[ \frac{A_{0\sigma}}
       {\omega+\mu_\sigma-\tilde{\Omega}_{\sigma}(\omega)}
       + \frac{A_{2\bar\sigma}}
       {\omega+\mu_\sigma-U-\tilde{\Omega}_{\sigma}(\omega)}
       \bigg]^{-1} \nonumber\\
& &{}  + \tilde{\Omega}_{\sigma}(\omega),
\end{eqnarray}}%
where
\begin{equation}
\tilde{\Omega}_{\sigma}(\omega)=J_\sigma(\omega)-
\frac{R_\sigma(\omega)}{A_{0\sigma}A_{2\bar\sigma}} +
\frac{V_\sigma
\varphi_\sigma(\omega)}{A_{0\sigma}A_{2\bar\sigma}},
\end{equation}
and
{\arraycolsep=-4pt
\begin{eqnarray}
  & & {R_\sigma(\omega)  =  -\frac{\langle X^{\sigma\sigma}+
  X^{\bar\sigma\bar\sigma}\rangle}{2}
  J_{\bar\sigma}(\omega+\mu_\sigma-\mu_{\bar\sigma})}  \nonumber\\
  & & -\frac{\langle X^{\sigma\sigma}-X^{\bar\sigma\bar\sigma}\rangle}{2\pi}
  \!\int\limits_{-\infty}^{+\infty}\! \frac{\rd \omega' \Imm J_{\bar\sigma}(\omega'+\ri\varepsilon)}
  {\omega-\omega'-\mu_{\bar\sigma}+\mu_\sigma}
  \tanh\frac{\beta\omega'}{2} \nonumber\\
  & &+\frac{\langle X^{0 0}+ X^{2 2}\rangle}{2}
  J_{\bar\sigma}(U-\mu_\sigma-\mu_{\bar\sigma}-\omega) \nonumber \\
  & & {-}\frac{\langle X^{00}-X^{22}\rangle}{2\pi}
  \!\int\limits_{-\infty}^{+\infty}\!\! \frac{\rd \omega' \Imm  J_{\bar\sigma}(-\omega'-\ri \varepsilon)}
  {\omega-\omega'+\mu_{\bar\sigma}+\mu_\sigma-U}
  \tanh\frac{\beta\omega'}{2}.
\end{eqnarray}}%

The approach used here for the approximate solution of the
single-site problem can be called as the generalized Hubbard-III
(GH3) approximation. It becomes the standard Hubbard-III
approximation in the case of the usual Hubbard model at
half-filling with spin degeneration
($n_\uparrow=n_\downarrow=1/2$, $\mu_\uparrow=\mu_\downarrow=U/2$,
$\langle X^{00}\rangle=\langle X^{22}\rangle$, $\langle
X^{\sigma\sigma}\rangle=\langle X^{\bar\sigma\bar\sigma}\rangle$):
 {\arraycolsep=0.5pt
\begin{eqnarray}
 \Xi_\sigma^{-1}(\omega)&=&\bigg[ \frac{1/2}
       {\omega+U/2-3J_{\sigma}(\omega)}
       + \frac{1/2}
       {\omega-U/2-3J_{\sigma}(\omega)}
       \bigg]^{-1} \nonumber\\
& &{}  + 3J_{\sigma}(\omega).
\end{eqnarray}}%

The function $R_\sigma(\omega)$ describes band forming for
particles of sort $\sigma$ by the motion of particles of another
sort $\bar{\sigma}$ (scattering processes). The neglect of this
contribution ($R_\sigma(\omega)=0$) gives MAA approximation. If we
put $R_\sigma(\omega)=0$ and $\varphi_{\sigma}=0$, the system is
described within the simple AA approximation.

In the limit of infinite repulsion $U$ the following solution of
the single-site problem is obtained \cite{Sta03}:
\begin{equation}
  G_\sigma (\omega)=\frac{1-n_{\bar\sigma}}{\omega+\mu_\sigma-\frac{V_\sigma\varphi_\sigma}
           {1-n_{\bar\sigma}}-J_\sigma(\omega)+\frac{R_\sigma(\omega)}{1-n_{\bar\sigma}}}\,,
  \label{ss1}
\end{equation}
{\setlength\arraycolsep{0pt}
\begin{eqnarray}
   \lefteqn{ R_\sigma(\omega)  = -\frac{n_\sigma+n_{\bar\sigma}}{2}
   J_{\bar\sigma}(\omega+\mu_\sigma-\mu_{\bar\sigma})}\nonumber\\
  & & -\frac{n_\sigma-n_{\bar\sigma}}{2\pi}
  \int\limits_{-\infty}^{+\infty} \frac{\Imm
  J_{\bar\sigma}(\omega'+\ri\varepsilon)\rd\omega'}{\omega-\omega'-\mu_{\bar\sigma}+\mu_\sigma}
  \tanh\frac{\beta\omega'}{2}\,.
  \label{ss6}
\end{eqnarray}}%
 The constant
${\varphi_{\sigma}=\langle\xi_{\bar{\sigma}}\anni
X^{\bar{\sigma}0}\rangle+ \zeta\langle X^{\sigma
2}\xi_{\bar{\sigma}}\crea \rangle}$ can be calculated using the
exact relation given in the next section. The average particle
concentrations are calculated using the imaginary part of the
Green's functions (density of states):
\begin{equation}
\rho_\sigma(\omega)=-\frac{1}{\pi}\Imm
G_\sigma(\omega+\ri\varepsilon).
\end{equation}

The self-consistency conditions (a set of equations (\ref{sys2})
-- (\ref{sys3})) relate the coherent potential $J_\sigma$ to the
Green's function $G_\sigma$. For the Bethe lattice with a
semielliptic density of states
\begin{equation}
\rho^{\mathrm{Bethe}}_{\sigma} (\varepsilon)=\frac{2}{\pi
W_{\sigma}^2} \sqrt{W_{\sigma}^2-\varepsilon^2}\,, \quad
|\varepsilon|<W_{\sigma}
\end{equation}
we have
\begin{equation}
J_{\sigma}(\omega)=\frac{W_{\sigma}^2}{4}\, G_{\sigma}(\omega).
\end{equation}

In the case of the Falicov-Kimball model the unperturbed bandwidth
is zero ($2 W_B=0$, $J_B(\omega)=0$) for localized particles, and
the approach gives the exact equation for the Green's function of
itinerant particles $G_A(\omega)$. The density of state
$\rho_A(\omega)$ on the Bethe lattice is nonzero for
${|\omega+\mu_A|<W_A\sqrt{1-n_B}}$:
\begin{equation}
\rho_A(\omega)= \frac{2}{\pi W_A^2} \sqrt{W_A^2
(1-n_B)-(\omega+\mu_A)^2}. \label{rho_A}
\end{equation}
In this case the equations (\ref{ss1}), (\ref{ss6}) give an
explicit approximate expression for the Green's function of
localized particles:
\begin{equation}
  G_B (\omega)=\frac{1-n_A}{\omega+\mu_B-\frac{V_B\varphi_B}
           {1-n_A}+\frac{R_B(\omega)}{1-n_A}}\,,
  \label{GB_FK}
\end{equation}
where%
{\setlength\arraycolsep{0pt}
\begin{eqnarray}
   \lefteqn{ R_B(\omega)  = -\frac{n_B+n_A}{2}
   J_A (\omega+\mu_B-\mu_A)} \nonumber\\
  & & -\frac{n_B-n_A}{2\pi}
  \int\limits_{-\infty}^{+\infty} \frac{\Imm
  J_A(\omega'+\ri\varepsilon)\rd\omega'}{\omega-\omega'-\mu_A+\mu_B}
  \tanh\frac{\beta\omega'}{2}\,.
  \label{RB_FK}
\end{eqnarray}}%

\section{The projecting coefficients in equations of motion
for Hubbard operators}

The average values $\langle X^{pq}\xi_{\sigma}\rangle$ are
calculated using corresponding Green's functions according to the
spectral theorem
\begin{equation}
\langle X^{pq}\xi_{\sigma}\anni \rangle =
\int\limits_{-\infty}^{+\infty}
{\frac{\rd\omega}{\re^{\beta\omega}+1}} \Big[ -2\Imm\rl
\xi_{\sigma}\anni|X^{pq}\rr_{\omega+\ri\varepsilon} \Big].
\end{equation}
These Green's functions can be calculated using the exact relation
(\ref{fi_ex01}) derived in Appendix
\begin{equation}
 V_{\sigma} \rl \xi_{\sigma}\anni|X^{pq}\rr_{\omega}
 =  J_{\sigma}(\omega) \rl
a_{\sigma}|X^{pq}\rr_{\omega}.
\end{equation}

For the asymmetric Hubbard model we have to calculate the
coefficients $V_{\sigma}\varphi_{\sigma}$:
\begin{equation}
\varphi_{\sigma}=\langle\xi_{\bar{\sigma}}X^{\bar{\sigma}0}\rangle+
\zeta\langle X^{\sigma 2}\xi_{\bar{\sigma}}\crea
\rangle=\langle\xi_{\bar{\sigma}} (X^{\bar{\sigma}0}+\zeta
X^{2\sigma} )\rangle
\end{equation}
which are expressed as
\begin{equation}
V_{\sigma}\varphi_{\sigma}=\int_{-\infty}^{+\infty}
\frac{\rd\omega}{\re^{\beta\omega}+1} \Imm
\rho_{\sigma}^{\varphi}(\omega+\ri\varepsilon)\,,
\end{equation}
{\arraycolsep=0.5pt
\begin{eqnarray}
\rho_{\sigma}^{\varphi}(\omega) &=& 2  J_{\bar{\sigma}}(\omega)
\rl X^{0\bar{\sigma}}-\zeta X^{\sigma 2} | X^{\bar{\sigma}0}+\zeta
X^{2\sigma} \rr_{\omega}  \nonumber\\
&=& 2  J_{\bar{\sigma}}(\omega) \big[
 \rl X^{0\bar{\sigma}}|X^{\bar{\sigma}0}  \rr_{\omega}-
\rl X^{{\sigma}2}|X^{2{\sigma}}  \rr_{\omega}  \big].
\end{eqnarray}}%
In the limit of infinite on-site repulsion $U$ the state with
double occupation is excluded and we have
\begin{equation}
\rho_{\sigma}^{\varphi}(\omega)=\frac{1}{\pi}
J_{\bar{\sigma}}(\omega) G_{\bar\sigma}(\omega)\,.
\label{fi_ex_rho}
\end{equation}%
For the Falicov-Kimball model it is possible to calculate the
exact Green's functions for itinerant particles and it gives the
exact expression for $V_{B}\varphi_{B}$.

Let us note that previously this parameter was calculated
approximately using the Green's functions obtained by means of the
linearized equations of motion and neglecting the irreducible
parts \cite{Sta00,Sta02,Sta03}. In the model with infinite $U$
this approximate value is
\begin{equation}
\rho_{B}^{\varphi}(\omega)=\frac{1}{\pi} J_{A}(\omega)
\frac{1-n_B}{\omega+\mu_B}\,, \label{fi_approx}
\end{equation}
and it corresponds to the approximation of the series
(\ref{fi_ser02}) where only the first term (the zero approximation
for $G_A$) is taken into account.

\begin{figure}%[htbp!]
\includegraphics[width=0.45\textwidth]{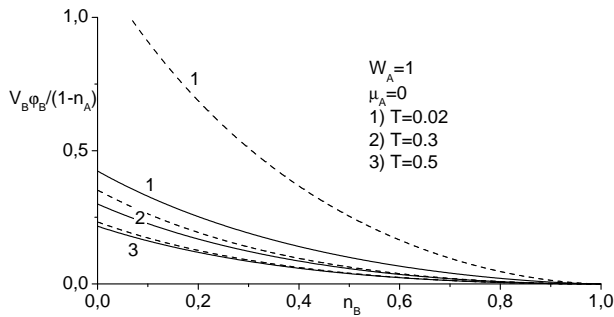}
\caption{\label{fi_comp} The exact values of $\varphi$ (solid
lines) are compared with the approximate ones calculated using the
relation (\ref{fi_approx}) (dashed lines) for various temperatures
and $n_B$. One can see that the approximation is applicable only
for very high temperatures.}
\end{figure}

There is shown a comparison between the approximate and the exact
values of $\varphi_B$ in \fig\ref{fi_comp}. The improvement given
by the exact relation (\ref{fi_ex_rho}) allows us to investigate
the system at low temperatures where summing up the whole series
(\ref{fi_ser02}) is essential.

\section{Results}
Dependences of chemical potentials $\mu_A$ and $\mu_B$  on the
particle concentrations are calculated using corresponding
densities of states. DOS of localized particles is obtained as an
imaginary part of the Green's function
\begin{equation}
 \rho_B (\omega)=-\frac{1}{\pi} \Imm G_B(\omega+\ri \varepsilon)\,.
\end{equation}
This Green's function has the correct analytic properties
(relations between imaginary and real parts)  within the
considered approximations (GH3, MAA, AA). The approximate DOS
always has the correct sign and the sum rule is fulfilled: the
integral of DOS over all frequency is equal to unity for finite
$U$ and is equal to $1-n_{\bar\sigma}$ for $U\rightarrow\infty$
when the upper band tends to infinity.

\begin{figure}
\includegraphics[width=0.43\textwidth]{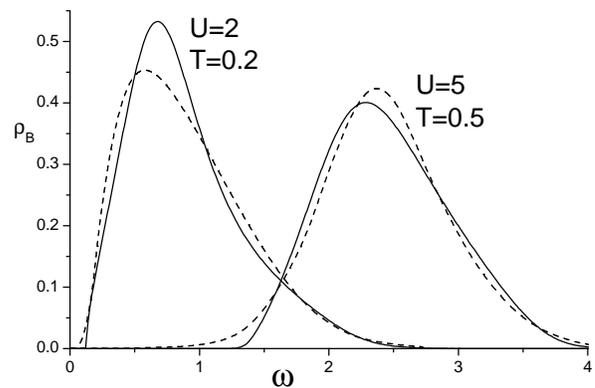}
\caption{\label{hf_cubic} DOS of localized particles for the
Falicov-Kimball model on the hypercubic lattice with
noninteracting DOS
$\rho_A^{\mathrm{hyp}}(\varepsilon)=\pi^{-1/2}\exp(-\varepsilon^2)$
at half-filling ($n_A=n_B=1/2$). Solid line -- our approximation;
dashed line -- exact result \cite{Bra92}.}
\end{figure}

For the Falicov-Kimball model the density of states of localized
particles can be calculated exactly \cite{Bra92,Zla01,Fre03},  but
numerical results were obtained mostly at half-filling. The
constant $\varphi_B$ is zero at half-filling ($n_A=n_B=1/2$)
because of the particle-hole symmetry. In this case we have the
simple approximate solution of the single-site problem
\begin{equation}
G_B (\omega)=\frac{\omega-2J_A(\omega)}{\omega^2-U^2/4-2\omega
J_A(\omega)}.
\end{equation}
This result at half-filling is independent of temperature, but for
high temperatures and large values of $U$ the approximate scheme
reproduces the exact results (\fig\ref{hf_cubic}).

For simplicity we restrict our investigation to the
Falicov-Kimball model with the infinite repulsion $U$ on the Bethe
lattice. In this case the model describes the system with an
average site occupation no more than unity. There is a homogeneous
state for temperatures larger than critical $T_{\mathrm{c}}\approx
0.060W_A$. For lower temperatures ($T<T_{\mathrm{c}}$) there are
various types of phase transitions depending on a thermodynamic
regime \cite{Fre99,Let99,Sta03}.

First, we consider a possibility of describing the phase
transitions using the approximate equations. For this reason, the
behavior of dependences of the chemical potential $\mu_B$ on the
concentration of localized particles $n_B$ is investigated. The
phase transition is indicated by the  thermodynamic unstable
region where $\partial \mu_B/\partial n_B<0$. When the critical
temperature is approached from below, the dependence $\mu_B(n_B)$
becomes monotonic ($\partial \mu_B/
\partial n_B\geq 0$).

\begin{figure*}%[htbp!]
\includegraphics[width=0.32\textwidth]{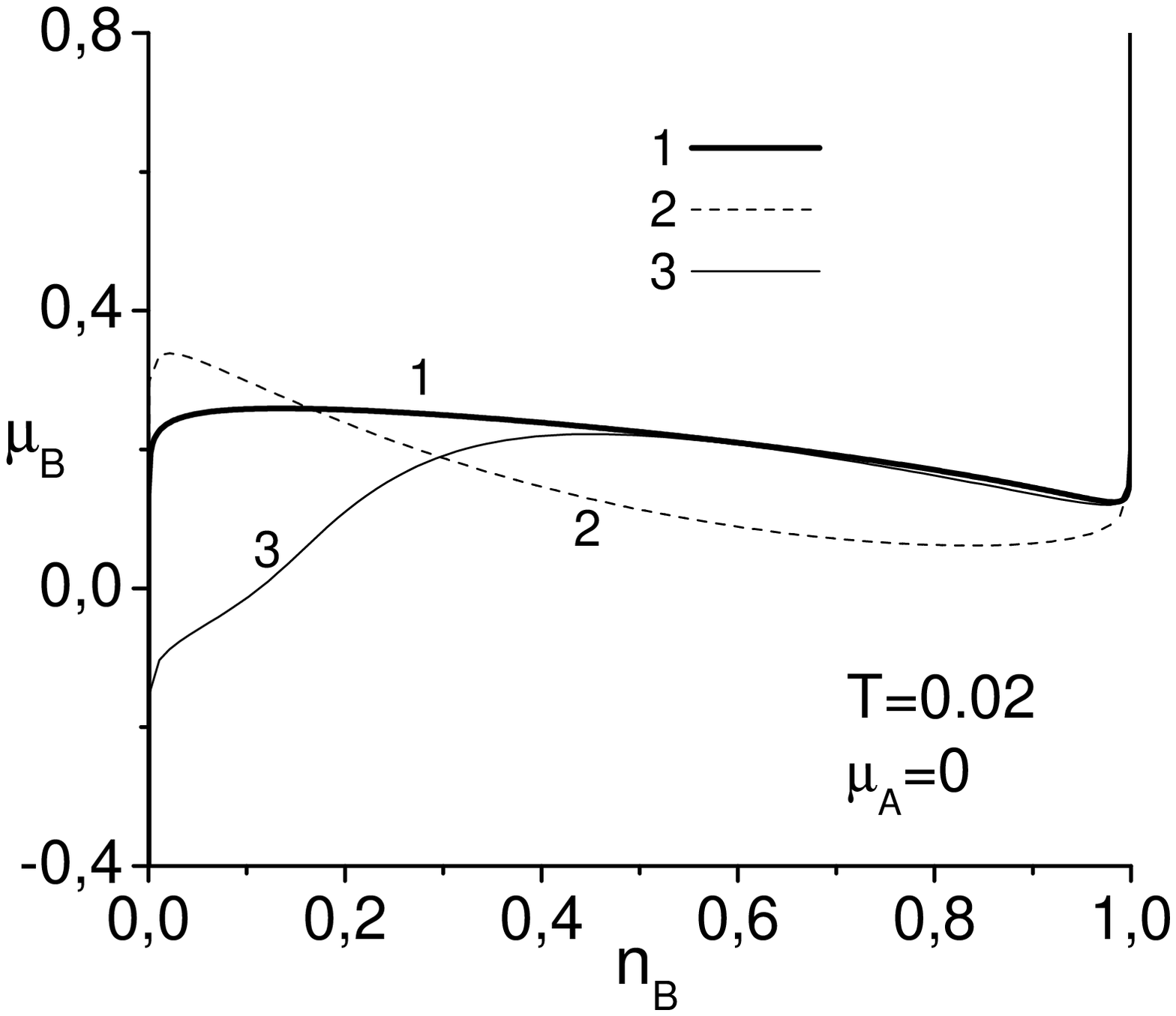}
\hfill
\includegraphics[width=0.32\textwidth]{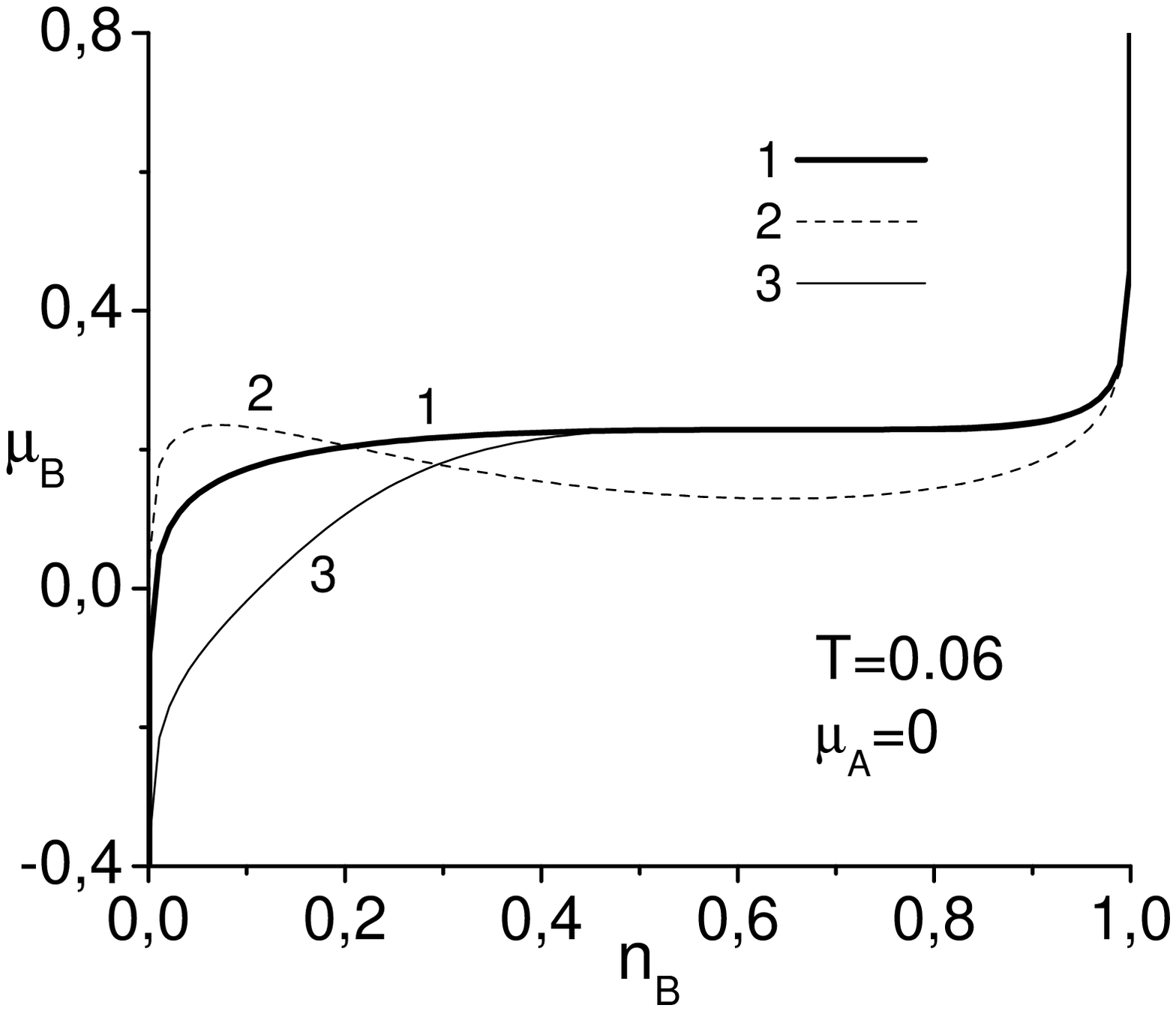}
\hfill
\includegraphics[width=0.32\textwidth]{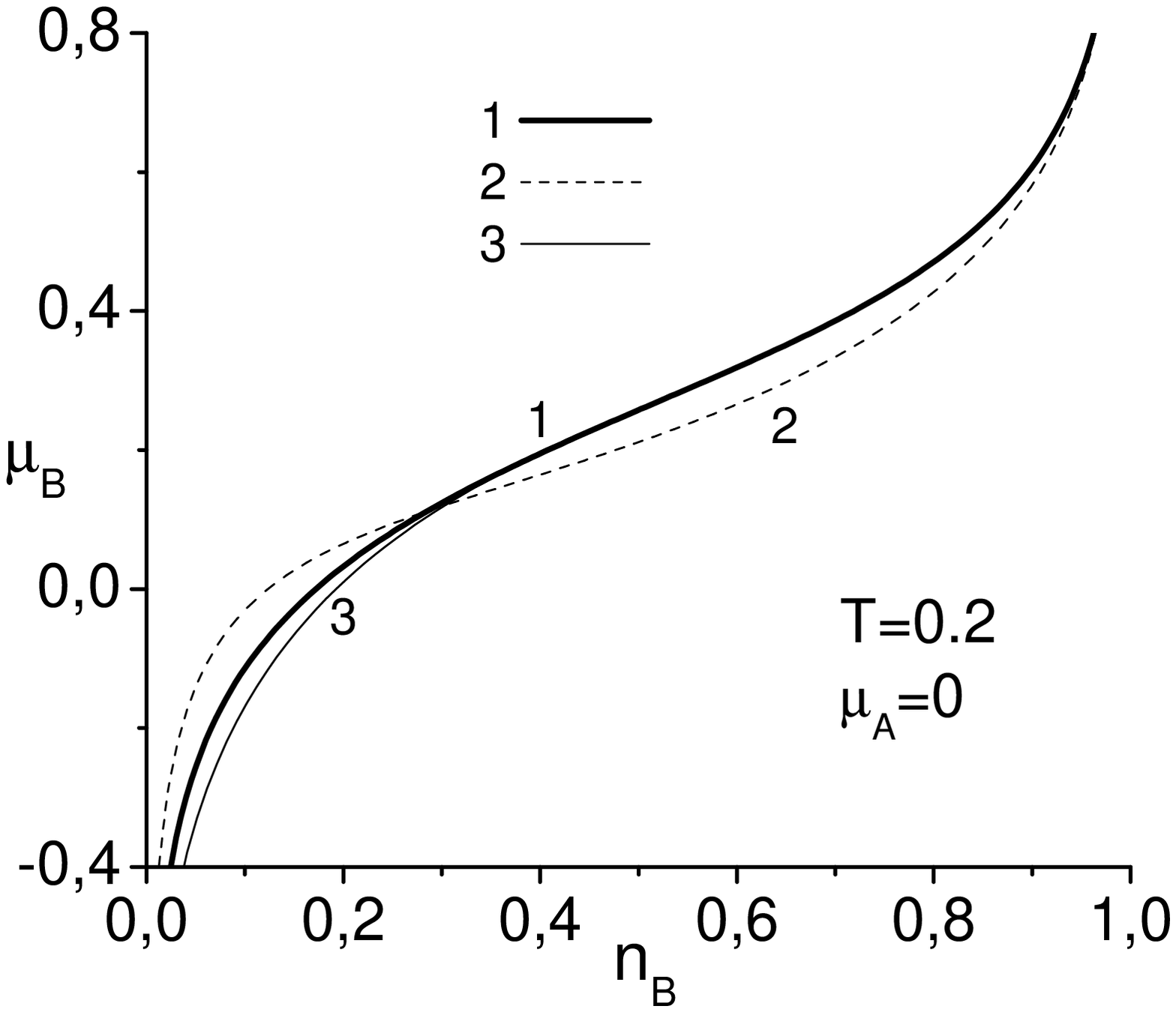}
\caption{\label{fig_mn_comp} The dependence of $\mu_B$ on $n_B$ in
different approximations is compared with the exact result
obtained thermodynamically. The parameter values: $W_A=1$,
$W_B=0$, $U=\infty$. 1 -- exact result; 2 -- MAA; 3 -- GH3.}
\end{figure*}

\begin{figure*}%[htbp!]
\includegraphics[width=0.32\textwidth]{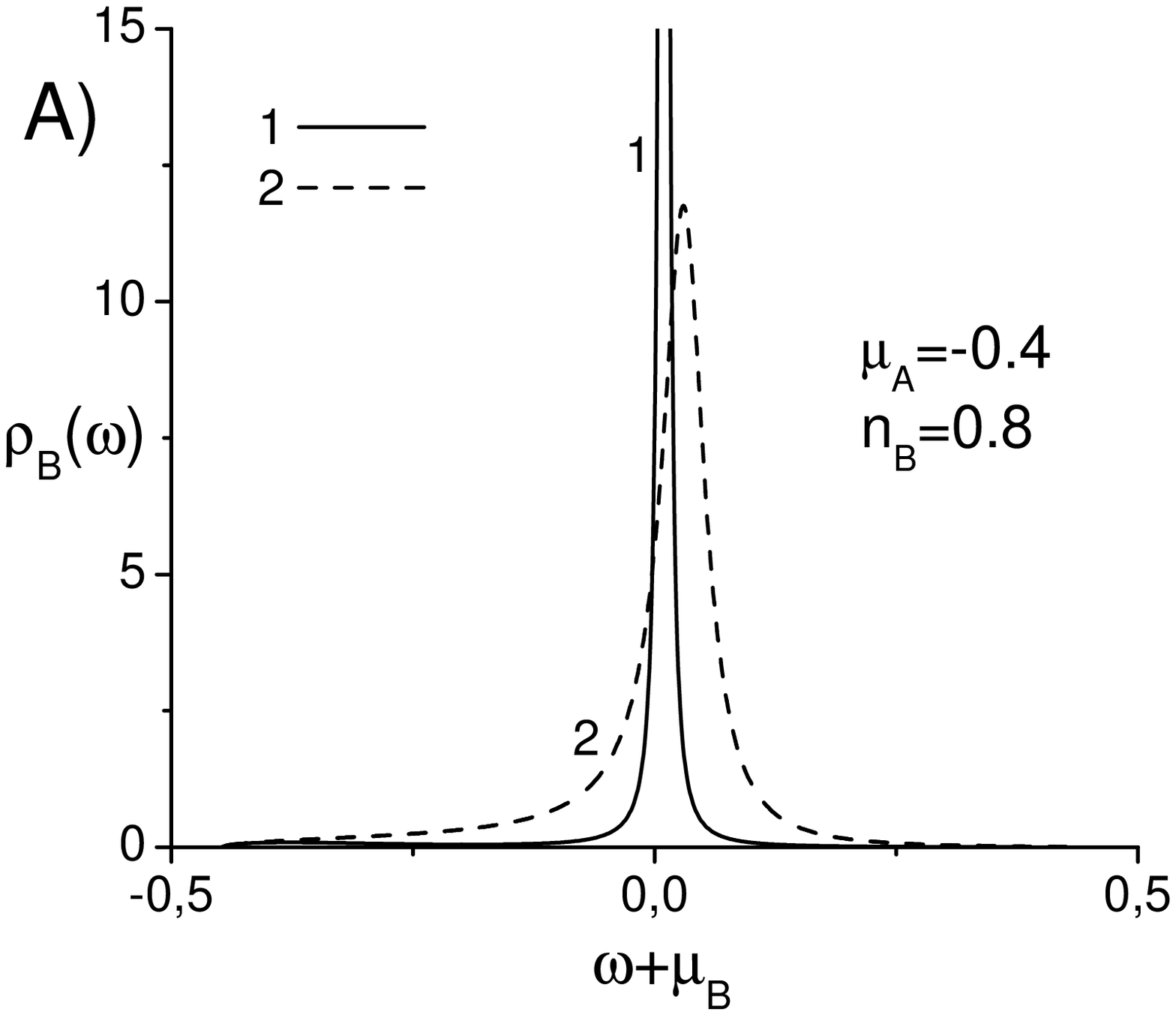}
\hfill
\includegraphics[width=0.32\textwidth]{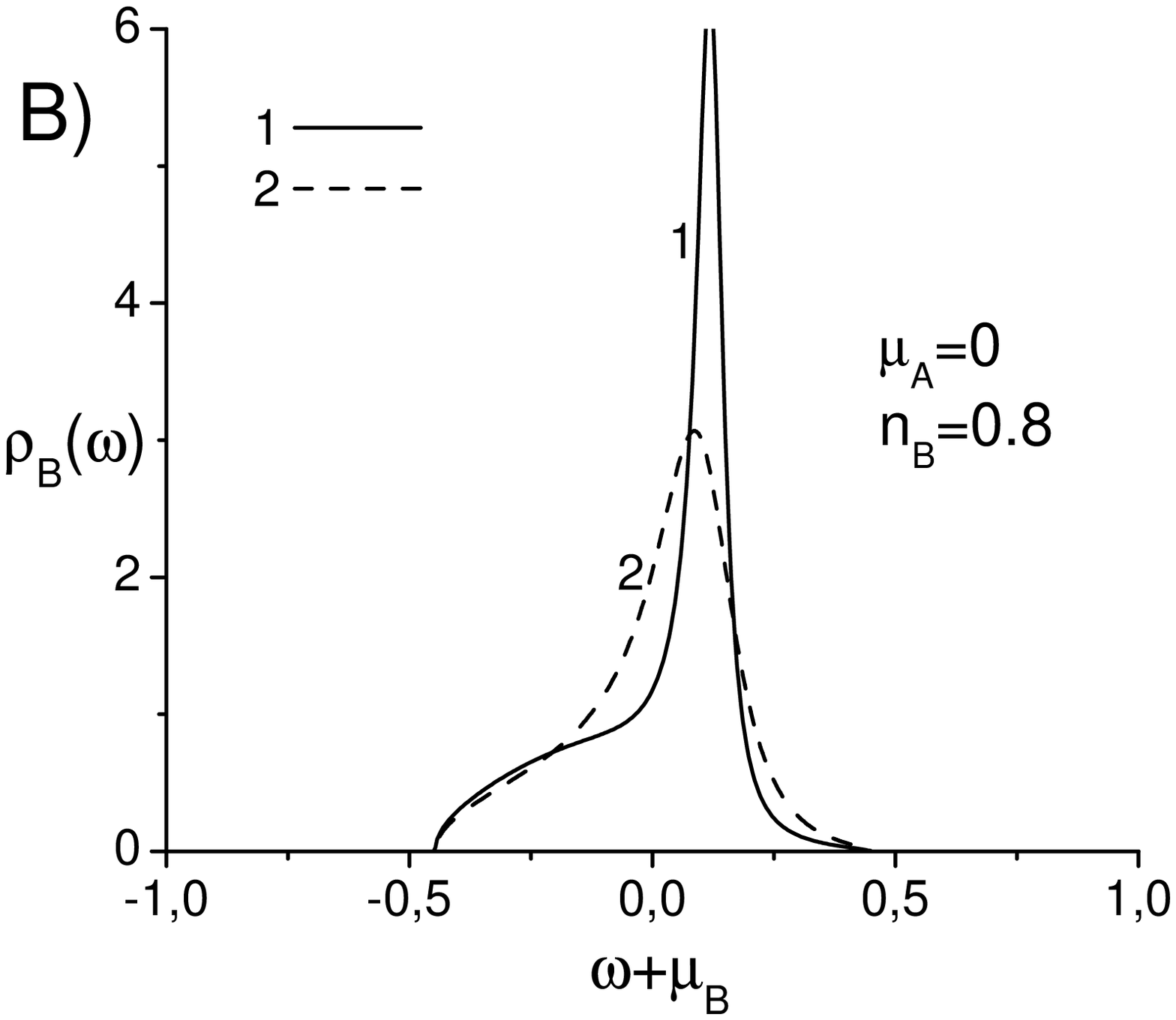}
\hfill
\includegraphics[width=0.32\textwidth]{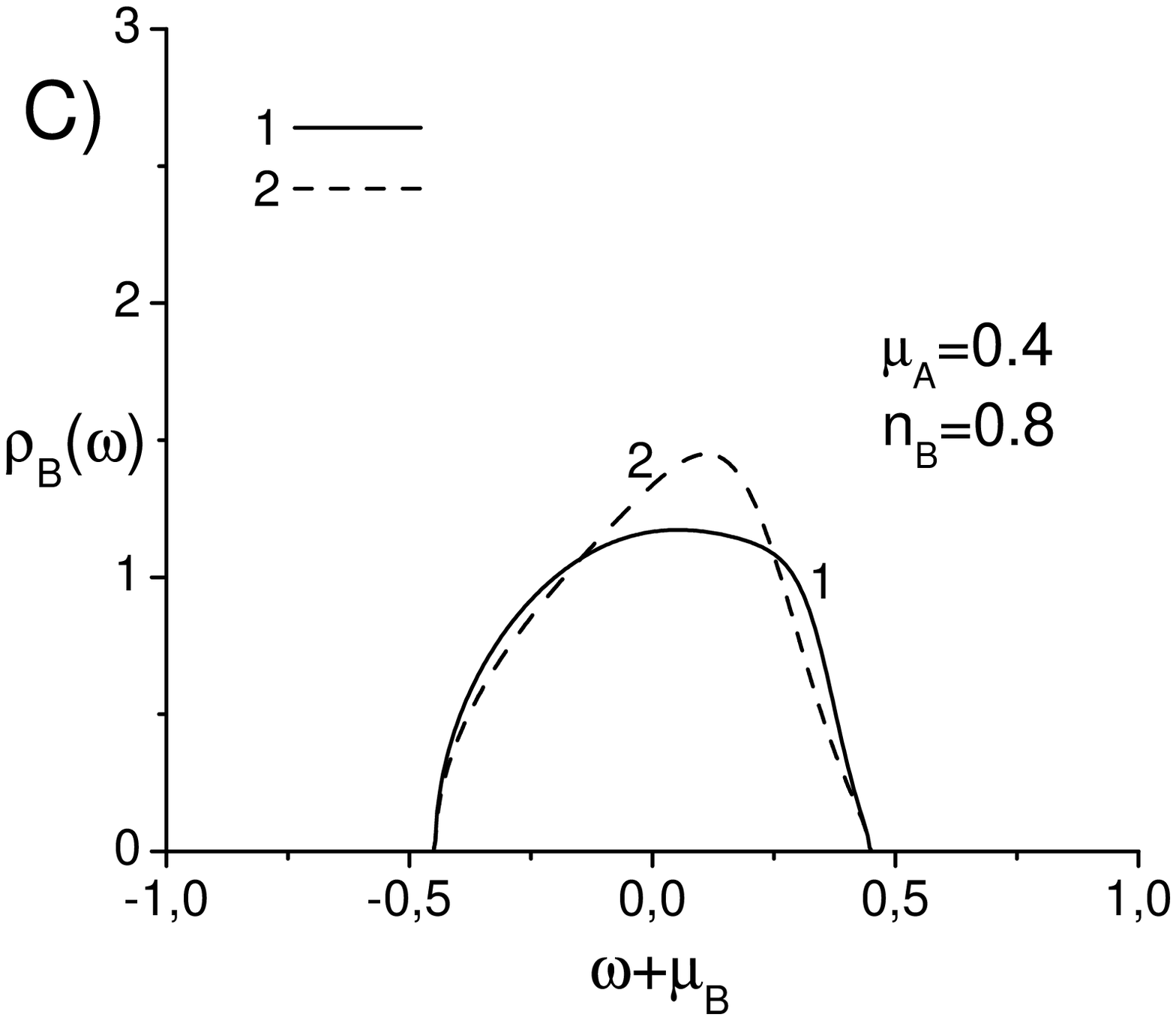}
\caption{\label{fig_dos} DOS of localized particles within the GH3
approximation for various $\mu_A$ and temperatures. $W_A=1$;
$W_B=0$; $U=\infty$; 1) $T=0.06$, 2) $T=0.2$.}
\end{figure*}

In \fig\ref{fig_mn_comp} the approximate curves  $\mu_B(n_B)$ are
compared with the exact results obtained thermodynamically
\cite{Sta02,Sta03}. The alloy-analogy based approximations give
the density of states of localized particles in a form of a
noninteracting delta-function, which is correct in the atomic
limit ${t_{A}=0}$. Thus, the MAA approximation can give reasonable
results only when the chemical potential $\mu_A<0$ (i.e. the
concentration of itinerant particles tends to zero). This
approximation shows the presence of a phase transition. However,
it overestimates the critical temperature. In
\fig\ref{fig_mn_comp} the exact curve at $T=0.060$ is already
monotonic but the MAA approximation shows the region with unstable
concentration values, i.e. there is still a phase transition.

The GH3 approximation, unlike the MAA approximation, incorporates
the scattering processes forming the energy band of localized
particles. This is crucial for the calculation of thermodynamic
quantities. The approximate curves coincide with the exact ones in
a wide range of temperatures for the concentrations larger than
some value which depends on $\mu_A$. In \fig\ref{fig_mn_comp}
($\mu_A=0$) one can see a good agreement with the exact result at
$\mu_A=0$ for concentrations $n_B\gtrsim 0.5$. At temperatures
lower than the critical one the approximation clearly indicates
the phase transition and gives the correct value for
$T_\mathrm{c}\backsimeq 0.060$.

Spectra of localized particles are plotted in  \fig\ref{fig_dos}.
In this case temperatures are higher than the critical one, so a
homogeneous state is stable. The parameter values are chosen so
that the approximation gives the correct thermodynamic
$\mu_B(n_B)$ relations. The energy band of itinerant particles (A)
depends only on a concentration $n_B$ and its width is
$2W_A\sqrt{1-n_B}$. However, the band of localized particles (B)
is generated by scattering processes and its spectral shape
depends on the concentration $n_B$, the chemical potential $\mu_A$
(or the corresponding concentration $n_A$) and temperature.

Depending on the values of the chemical potential of itinerant
particles $\mu_A$ (or $n_A$) there are two limiting cases with
different properties of the localized particle spectrum. For very
small $n_A$ (negative $\mu_A$) the system is close by its behavior
to the atomic limit, i.e., the spectrum $\rho_B(\omega)$ is in the
form of a delta peak. The sharp peak is slightly broadened by the
scattering of itinerant particles (\fig\ref{fig_dos}A,
$\mu_A=-0.4$). The contrary case with the nearly filled bands is
when the total concentration of particles tends to unity,
$n_A+n_B\rightarrow 1$ (\fig\ref{fig_dos}C, $\mu_A=0.4$). In this
case the peak vanishes because the contribution of the coherent
potential of mobile particles $J_A(\omega)$ and the
cor\-re\-spond\-ing function $R_B(\omega)$ (\ref{RB_FK}) becomes
larger and the simple pole in (\ref{GB_FK}) disappears. The
spectrum $\rho_B(\omega)$ corresponds by the form to the lower
Hubbard subband (or the lower subband for the half-filled
Falicov-Kimball model obtained in Ref.~\onlinecite{Bra92}) with
the chemical potential located in a gap  in the strong coupling
limit. The intermediate case with the broad band superimposed by
the sharp peak is shown in \fig\ref{fig_dos}B.

\begin{figure}
%\parbox{0.32\textwidth}
{\includegraphics[width=0.24\textwidth]{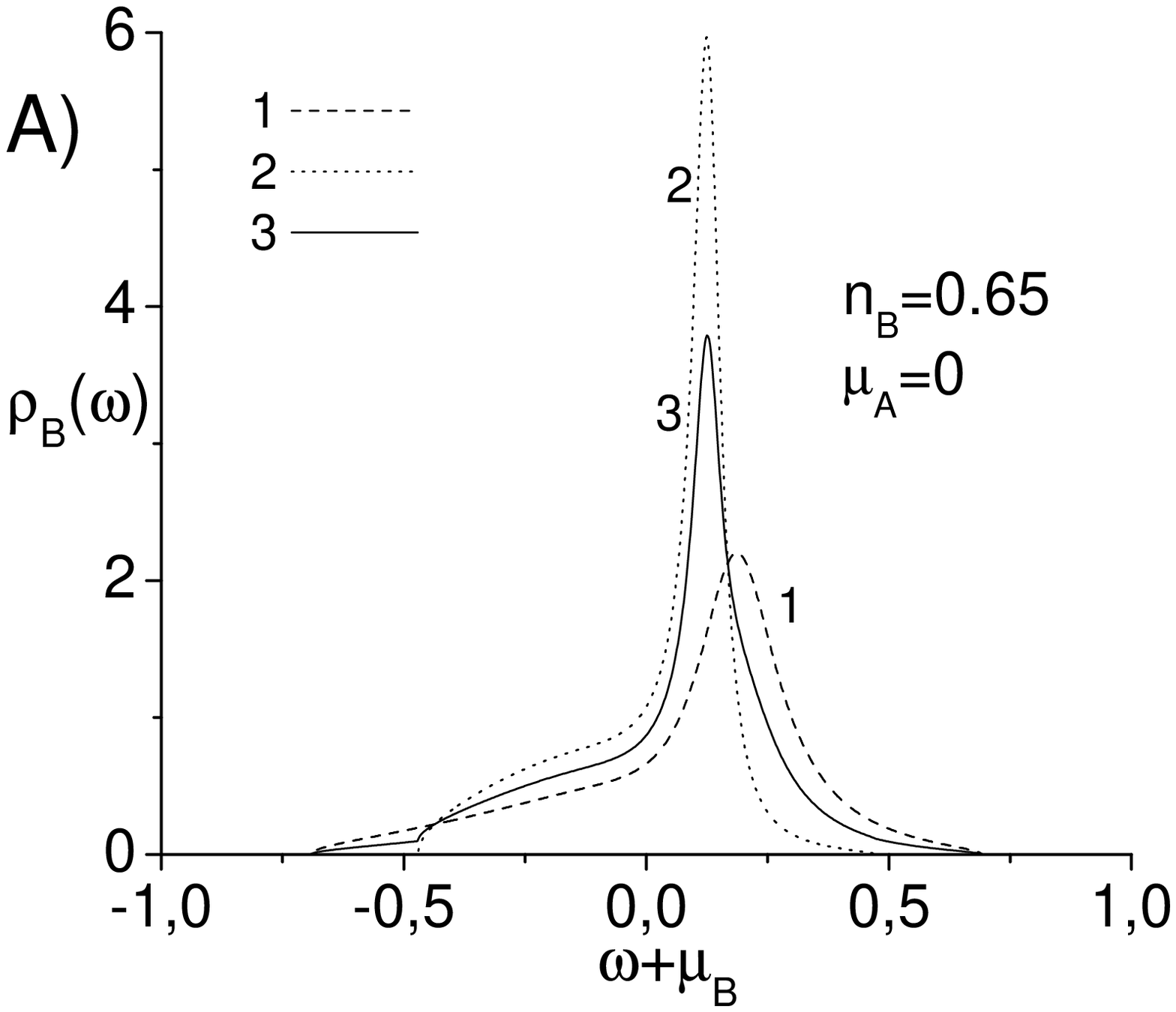}}% %\hfill
%\parbox{0.32\textwidth}
{\includegraphics[width=0.24\textwidth]{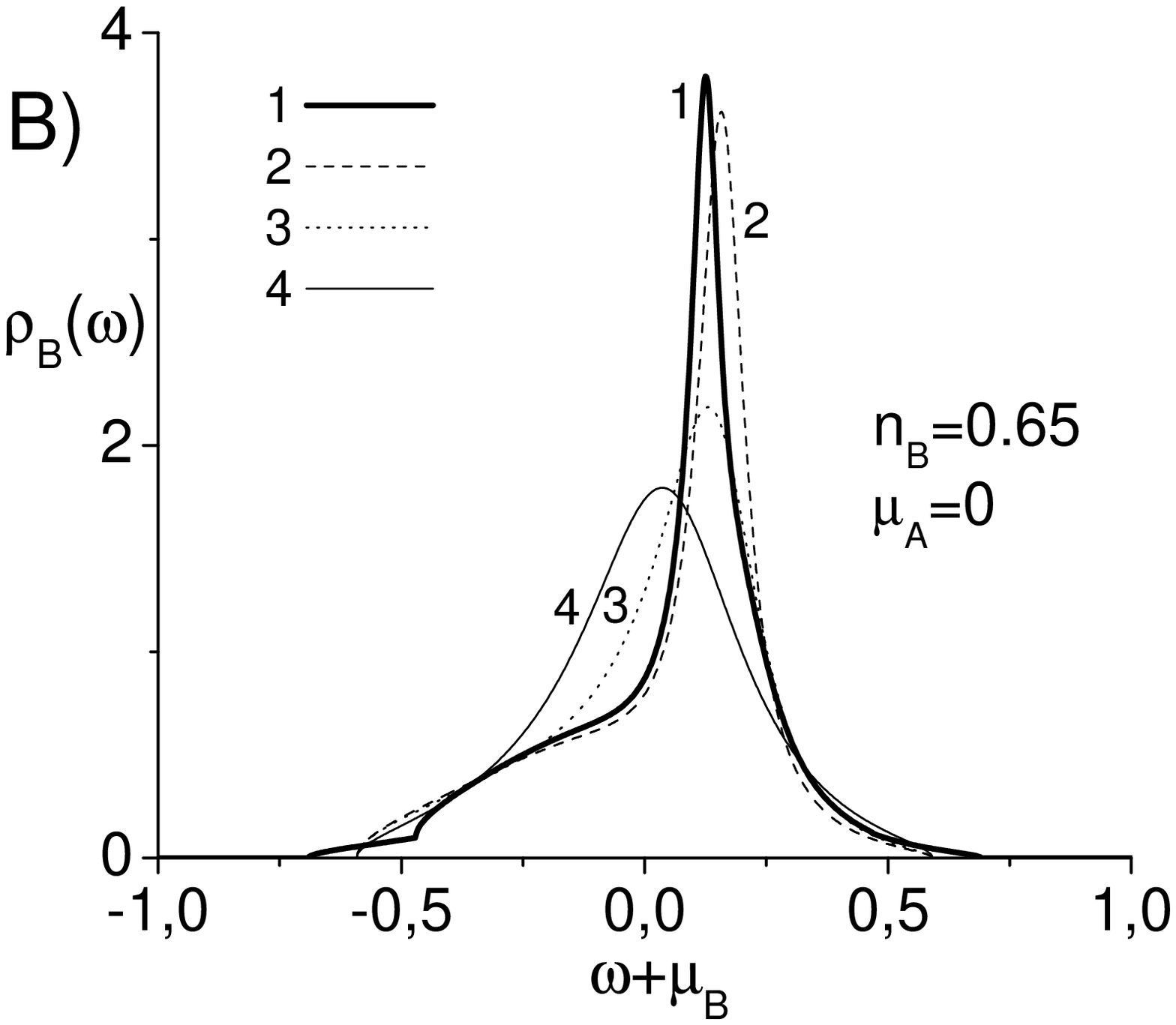}} %\hfill
%\parbox{0.35\textwidth}
{\caption{\label{fig_dos_sep} DOS of localized particles within
the GH3 approximation ($W_A=1$, $W_B=0$). A) At temperatures below
$T_{\mathrm{c}}$ ($T=0.059$) the state is separated into two
phases with different particle concentrations (1,2 -- the spectra
of each component; 3 -- the superposition of 1 and 2). B) DOS for
various
 temperatures  (1 -- $T=0.059$, 2 -- $T=0.061$, 3 -- $T=0.2$,
4 -- $T=1.0$).}}
\end{figure}

States with concentration values in some region  (where $\partial
\mu_{\sigma}/\partial n_{\sigma}<0$) are thermodynamically
unstable at low temperatures. So, the presence of phase
transitions should be taken into consideration. There are phase
transitions between homogeneous phases with a concentration jump
in the thermodynamic regime with the fixed chemical potentials
($\mu_A=\const$, $\mu_B=\const$). In this case the density of
states changes instantly as the concentration jumps. If one of the
concentrations ($n_A$ or $n_B$) is fixed the homogeneous state is
unstable and the phase separation takes place. The phase
transitions for the Falicov-Kimball model were investigated in
many works (see Ref.~\onlinecite{Fre03}); the regimes mentioned
above were investigated for the Falicov-Kimball model using the
exact thermodynamic equations in Ref.~\onlinecite{Sta03}.

Let us consider the thermodynamic regime with fixed values of
$\mu_A$ and $n_B$. In the homogeneous state the spectral function
of itinerant particles $\rho_A$ depends on $n_B$ and $\mu_A$
(\req\ref{rho_A}) and is independent of temperature. However, in
the phase separated state the spectrum of the whole system can be
considered as a superposition of spectra of each component. The
homogeneous state is unstable for the concentration values $n_B
\in (n_{B1},n_{B2})$ and the system is separated into two
different phases with the concentrations of $B$ particles
$n_{B1}\approx 0.52$ and $n_{B2}\approx 0.78$ at $T=0.059W_A$. The
concentrations of the components ($n_{B1}$, $n_{B2}$) depend on
temperature \cite{Sta03}. Thus, the spectra $\rho_A$ and $\rho_B$
are temperature dependent in the segregated phase. In
\fig\ref{fig_dos_sep}(A) DOS of localized particles is plotted at
temperature $T=0.059W_A<T_{\mathrm{c}}$. In
\fig\ref{fig_dos_sep}(B) the spectrum at various temperatures is
compared. The bandwidth is larger in the segregated state than in
the homogeneous state ($T>T_{\mathrm{c}}$) and is temperature
dependent.

\section{Conclusions}

The approximate analytic approach within DFMT for calculating the
single-particle Green's functions of the asymmetric Hubbard model
is developed and improved. This approximation allows one to
investigate the model for various concentration values. The method
is tested on the Falicov-Kimball model with the infinite on-site
repulsion. It is shown that for high enough temperatures or large
con\-cen\-tra\-tions of localized particles the approximate
approach reproduces exact values of chemical potentials. The
approximation can correctly indicate the instability of a
homogenous state and the presence of phase transitions.

The generalized Hubbard-III approximation (GH3) partially includes
the scattering of particles into the theory and describes the
formation of the band of localized particles. In the infinite-$U$
limit the spectrum of localized particles is obtained for various
particle concentrations and temperatures. The form of this
spectrum continuously changes from a delta peak to the
characteristic form of the lower subband of the spectrum in the
Hubbad-III approximation when the chemical potential of itinerant
particles increases.

\begin{acknowledgments}
This work was partially supported by the Fundamental Researches
Fund of the Ministry of Ukraine of Science and Education (Project
No. 02.07/266).

The authors thank Dr. A.~Shvaika for helpful discussions.
\end{acknowledgments}

\appendix*
\section{Exact relations between Green's functions}

%\begin{theorem}
\emph{Let us consider the effective single-site problem in terms
of the auxiliary Fermi-field
\begin{equation}
\hat{H}_{\mathrm{eff}}=\hat{H}_0+\hat{H}_{\xi}+\sum_{\sigma=1}^{m}
V_{\sigma}\Big(a_{\sigma}\crea \xi_{\sigma}\anni\!
+\xi_{\sigma}\crea a_{\sigma}\anni\! \Big),
\end{equation}
$\hat{H}_0$ is a single-site Hamiltonian; $\hat{H}_{\xi}$ is an
auxiliary environment Hamiltonian. The number $m$ of sorts of
itinerant particles can be arbitrary ($m=1,\,2,\,3,\, \ldots$).
So, the effective Hamiltonian can describe the Falicov-Kimball
model ($m=1$) and the Hubbard model ($m=2$). The algebra of $\xi$
operators is defined by the anticommutation relations
\begin{equation}
\{\xi_{\alpha},\,\xi_{\beta}\}=\{\xi_{\alpha}\crea , \,
\xi_{\beta}\crea \}=0, \quad \{\xi_{\alpha}\crea , \,
\xi_{\beta}\anni\!\}=\delta_{\alpha,\beta}\,.
\end{equation}
It can be proved that the following relations take place
{\arraycolsep=1pt
\begin{eqnarray} V_{\sigma}\langle\langle
\xi_{\sigma}|\hat{A}\rangle\rangle_{\omega}&=& J_{\sigma}(\omega)
\langle\langle a_{\sigma}|\hat{A}\rangle\rangle_{\omega},
\label{fi_ex01}\\[0.5ex]
2\pi V_{\sigma}^2 \langle\langle
\xi_{\sigma}\anni|\xi_{\sigma}\crea \rangle\rangle_{\omega} &=&
J_{\sigma}(\omega)+2\pi J^2_{\sigma}(\omega) \langle\langle
a_{\sigma}\anni | a_{\sigma}\crea \rangle\rangle_{\omega},
\label{fi_ex02}
\end{eqnarray}}%\\[0ex]
where $\hat{A}$ is the arbitrary Fermi-operator that
anticommutates with the $\xi$ operators, and $J_{\sigma}(\omega)$
is the Green's function for the unperturbed Hamiltonian
$$J_{\sigma}(\omega)=2\pi V^2_{\sigma}
\langle\langle\xi_{\sigma}\anni |\xi_{\sigma}\crea
\rangle\rangle_{\omega}^{\xi}.$$}%
%\end{theorem}

\emph{Proof.}
 Thermodynamic perturbation theory can be formulated
based on the interaction representation for the statistical
operator with the use of the $H_0+H_{\xi}$ operator as the
zero-order Hamiltonian:
\begin{eqnarray}
&&{}\hat{\rho}=\re^{-\beta H_{\reff}}= \re^{-\beta(H_0+H_{\xi})}
\hat{\sigma}(\beta),\\
&&{}\hat{\sigma}(\beta)=\re^{-\beta(H_0+H_{\xi})}\mathcal{T}_{\tau}
\exp \bigg[-\int_0^{\beta}\!\!\rd\tau \hat{H}_{\mathrm{int}}
(\tau)\bigg].
\end{eqnarray}%
The part of the interaction Hamiltonian describing one sort
($\sigma$) of particles is separated off:
\begin{equation}
\hat{H}_{\mathrm{int}}=V_{\sigma}\Big(a_{\sigma}\crea
\xi_{\sigma}\anni + \xi_{\sigma}\crea a_{\sigma}\anni \Big) +
\hat{B}_{\sigma}.
\end{equation}
The residue $\hat{B}_{\sigma}$ commutates with the operators of
the chosen sort $\sigma$:
$[\hat{B}_{\sigma},\,\xi_{\sigma}]=[\hat{B}_{\sigma},\,\xi_{\sigma}\crea]=0$.
Thus, at the perturbation expansion the operator
$\xi_{\sigma}^{(\dagger)}$ does not have to be paired with the
operator $\hat{B}_{\sigma}$. Here we introduce notations for the
Green's functions {\arraycolsep=1pt
\begin{displaymath}
\begin{array}{lll}
\mathcal{G}_{\sigma}(\tau-\tau') & = & \langle
\mathcal{T}_{\tau}\xi_{\sigma}\crea (\tau)\xi_{\sigma}\anni(\tau')\rangle_0\\
 G_{\sigma}(\tau-\tau') & = & \langle
\mathcal{T}_{\tau}a_{\sigma}\crea (\tau)a_{\sigma}\anni(\tau')\rangle\\
\Phi_{\sigma}(\tau-\tau') & = & \langle
\mathcal{T}_{\tau}\hat{A}_{\sigma}(\tau)\xi_{\sigma}(\tau')\rangle\\
G^A_{\sigma}(\tau-\tau') & = & \langle
\mathcal{T}_{\tau}\hat{A}_{\sigma}(\tau)a_{\sigma}(\tau')\rangle
\end{array}
\end{displaymath}}%\\
The perturbation theory expansion of the scattering matrix
$\hat{\sigma}(\beta)$  gives the following series for the Green's
function $\Phi_{\sigma}(\tau-\tau')$:
\begin{widetext}
\arraycolsep=-0.5ex
\begin{eqnarray}
&&{}
\Phi_{\sigma}(\tau-\tau')=\frac{V_{\sigma}^{2l+1}}{\langle\sigma(\beta)\rangle_0}
\sum_{p=0}^{\infty}\frac{(-1)}{(2p+1)!}\sum_{l=0}^{p}
\frac{(2p+1)!}{l!(l+1)!(2p-2l)!} \int_0^{\beta}\rd \tau_1 \ldots
\rd \tau_{2p+1} \nonumber\\[-1ex]
%\unitlength=2.27ex
\unitlength=1em
\put(4.15,1.1){\line(0,-1){0.25}\vector(1,0){6.0}\line(1,0){5.2}\line(0,-1){0.25}}
\put(4.05,1.3){\line(0,-1){0.5}\vector(1,0){8.9}\line(1,0){8.2}\line(0,-1){0.5}}
&&{} \times \langle\mathcal{T}_{\tau} \hat{A}(\tau)
\underline{\xi_{\sigma}(\tau')} \big(a_{\sigma}\crea
\xi_{\sigma}\anni \big)_{1}\cdots\big(a_{\sigma}\crea
\xi_{\sigma}\anni \big)_{l} \big(\xi_{\sigma}\crea a_{\sigma}\anni
\big)_{l+1}\cdots\big(\xi_{\sigma}\crea a_{\sigma}\anni
\big)_{2l+1} \hat{B}_{\sigma}(\tau_{2l+2})\cdots
\hat{B}_{\sigma}(\tau_{2p+1}) \rangle_0\,. \label{fi_ser01}
\end{eqnarray}%
The averaging of $\mathcal{T}$ products is performed in the
zero-order Hamiltonian according to the Wick's theorem by the
consecutive pairing. We start the pairing procedure from the
operator $\xi_{\sigma}\anni(\tau')$ and it is performed only with
the operators $\xi_{\sigma}\crea$. After the first pairing we have
the following expression:
\begin{eqnarray}
&&{} \Phi_{\sigma}(\tau-\tau')
=\frac{V_{\sigma}^{2l+1}}{\langle\sigma(\beta)\rangle_0}
\int_0^{\beta}\rd\tau''\mathcal{G}(\tau''-\tau')
\sum_{p=0}^{\infty}\sum_{l=0}^{p}
\frac{1}{l!l!(2p-2l)!} \int_0^{\beta}\rd \tau_1 \ldots \rd \tau_{2p} \nonumber\\
&&{} \times \langle\mathcal{T}_{\tau}
\hat{A}(\tau)a_{\sigma}(\tau'') \big(a_{\sigma}\crea
\xi_{\sigma}\anni \big)_{1} \cdots \big(a_{\sigma}\crea
\xi_{\sigma}\anni \big)_{l} \big( \xi_{\sigma}\crea
a_{\sigma}\anni \big)_{l+1} \cdots \big(\xi_{\sigma}\crea
a_{\sigma}\anni \big)_{2l} \hat{B}_{\sigma}(\tau_{2l+1})\cdots
\hat{B}_{\sigma}(\tau_{2p}) \rangle_0\,. \label{fi_ser02}
\end{eqnarray}%
\end{widetext}
Summing up the series in (\ref{fi_ser02}) gives the Green's
function ${G_{\sigma}^A(\tau-\tau')}$ and we have
\begin{equation}
\Phi_{\sigma}(\tau-\tau')=V_{\sigma}\int_0^{\beta}\rd \tau''
G_{\sigma}^A(\tau-\tau'')\mathcal{G}_{\sigma}(\tau''-\tau')\,,
\end{equation}
or in the Matsubara frequency representation:
\begin{equation}
\Phi_{\sigma}(\omega_n)=V_{\sigma}
G_{\sigma}^A(\omega_n)\mathcal{G}_{\sigma}(\omega_n)\,.
\end{equation}
Finally, in order to obtain the  expression (\ref{fi_ex01}), an
analytical continuation of the Green's functions from the
imaginary axis to the real one should be done
($\ri\omega_n\rightarrow{\omega+\ri \varepsilon}$,
$G_{\sigma}^A(\omega_n) \rightarrow 2\pi\langle\langle
a_{\sigma}|\hat{A}\rangle\rangle_{\omega}$,
$\Phi_{\sigma}(\omega_n) \rightarrow 2\pi V_{\sigma}\langle\langle
\xi_{\sigma}|\hat{A}\rangle\rangle_{\omega}$, \dots). If we put
$\hat{A}=\xi_{\sigma}\crea$, the pairing of the operator
$\xi_{\sigma}\anni(\tau')$ with $\xi_{\sigma}\crea(\tau)$ gives
the first term in (\ref{fi_ex02}). The second term is obtained
using the procedure the same as at deriving the relation
(\ref{fi_ex01}).

\end{document}